\documentstyle[12pt]{article}
\catcode`\@=11

\@addtoreset{equation}{section}
\def\theequation{\arabic{section}.\arabic{equation}}

\catcode`\@=11

\def\section{\@startsection{section}{1}{\z@}{3.5ex plus 1ex minus
   .2ex}{2.3ex plus .2ex}{\large\bf}}

\def\eqnarray{\stepcounter{equation}\let\@currentlabel=\theequation
    \global\@eqnswtrue
    \global\@eqcnt\z@\tabskip\@centering\let\\=\@eqncr
    $$\halign to \displaywidth\bgroup\@eqnsel\hskip\@centering
      $\displaystyle\tabskip\z@{##}$&\global\@eqcnt\@ne
       \hfil${{}##{}}$\hfil
      &\global\@eqcnt\tw@ $\displaystyle\tabskip\z@{##}$\hfil
       \tabskip\@centering&\llap{##}\tabskip\z@\cr}
\def\lefteqn#1{\hbox to 4\arraycolsep{$\displaystyle #1$\hss}}
\def\thesection{\arabic{section}.}

\def\appendix{\setcounter{section}{0}
        \def\thesection{Appendix.}
        \def\theequation{\Alph{section}.\arabic{equation}}}

\long\def\@makefntext#1{\parindent 0cm\noindent
\hbox to 1em{\hss$^{\@thefnmark}$}#1}
\def\IR{{\hbox{{\rm I}\kern-.2em\hbox{\rm R}}}}
\def\IH{{\hbox{{\rm I}\kern-.2em\hbox{\rm H}}}}
\def\IC{{\ \hbox{{\rm I}\kern-.6em\hbox{\bf C}}}}
\def\IZ{{\hbox{{\rm Z}\kern-.4em\hbox{\rm Z}}}}
\def\rref#1{(\ref{#1})}
\newcommand{\beq}{\begin{equation}}
\newcommand{\eeq}{\end{equation}}
\begin{document}
%
%
%
%
\def\citen#1{%
\edef\@tempa{\@ignspaftercomma,#1, \@end, }
\edef\@tempa{\expandafter\@ignendcommas\@tempa\@end}%
\if@filesw \immediate \write \@auxout {\string \citation {\@tempa}}\fi
\@tempcntb\m@ne \let\@h@ld\relax \let\@citea\@empty
\@for \@citeb:=\@tempa\do {\@cmpresscites}%
\@h@ld}
%
\def\@ignspaftercomma#1, {\ifx\@end#1\@empty\else
   #1,\expandafter\@ignspaftercomma\fi}
\def\@ignendcommas,#1,\@end{#1}
%
%
\def\@cmpresscites{%
 \expandafter\let \expandafter\@B@citeB \csname b@\@citeb \endcsname
 \ifx\@B@citeB\relax 
    \@h@ld\@citea\@tempcntb\m@ne{\bf ?}%
    \@warning {Citation `\@citeb ' on page \thepage \space undefined}%
 \else
    \@tempcnta\@tempcntb \advance\@tempcnta\@ne
    \setbox\z@\hbox\bgroup 
    \ifnum\z@<0\@B@citeB \relax
       \egroup \@tempcntb\@B@citeB \relax
       \else \egroup \@tempcntb\m@ne \fi
    \ifnum\@tempcnta=\@tempcntb 
       \ifx\@h@ld\relax 
          \edef \@h@ld{\@citea\@B@citeB}%
       \else 
          \edef\@h@ld{\hbox{--}\penalty\@highpenalty \@B@citeB}%
       \fi
    \else   
       \@h@ld \@citea \@B@citeB \let\@h@ld\relax
 \fi\fi%
 \let\@citea\@citepunct
}
%
\def\@citepunct{,\penalty\@highpenalty\hskip.13em plus.1em minus.1em}
%
%
\def\@citex[#1]#2{\@cite{\citen{#2}}{#1}}%
%
%
\def\@cite#1#2{\leavevmode\unskip
  \ifnum\lastpenalty=\z@ \penalty\@highpenalty \fi 
  \ [{\multiply\@highpenalty 3 #1
      \if@tempswa,\penalty\@highpenalty\ #2\fi 
    }]\spacefactor\@m}
\let\nocitecount\relax  
%
\begin{titlepage}
\vspace{.5in}
\begin{flushright}
UCD-92-23\\                             
gr-qc/9209011\\                         
September 1992\\
\end{flushright}
\vspace{.5in}
\begin{center}
{\Large\bf
The Modular Group, Operator Ordering, and Time\\
in (2+1)-dimensional Gravity}\\
\vspace{.4in}
{S.~C{\sc arlip}\footnote{\it email: carlip@dirac.ucdavis.edu}\\
       {\small\it Department of Physics}\\
       {\small\it University of California}\\
       {\small\it Davis, CA 95616}\\{\small\it USA}}
\end{center}

\vspace{.5in}
\begin{center}
{\large\bf Abstract}
\end{center}
{\small
A choice of time-slicing in classical general relativity permits the
construction of time-dependent wave functions in the ``frozen time''
Chern-Simons formulation of $(2+1)$-dimensional quantum gravity.  Because
of operator ordering ambiguities, however, these wave functions are not
unique.  It is shown that when space has the topology of a torus,
suitable operator orderings give rise to wave functions that transform
under the modular group as automorphic functions of arbitrary weights,
with dynamics determined by the corresponding Maass Laplacians on
moduli space.}
\end{titlepage}
\addtocounter{footnote}{-1}

Despite decades of research, physicists have not yet managed to
construct a workable quantum theory of gravity.  The considerable
effort that has gone into this venture has not been wasted, however:
we have gained a much better insight into the questions that must
be addressed, and we now know many of the ingredients that are likely
to be important in the final theory.  The purpose of this Letter is to
bring together three such fragments --- the ``problem of time,'' operator
orderings, and the mapping class group --- in the simplified context of
$(2+1)$-dimensional gravity.

The ``problem of time'' in quantum gravity appears in many guises, but
it takes its sharpest form in ``frozen time'' formulations such as
Chern-Simons quantization in $2+1$ dimensions.  Translations in
(coordinate) time are diffeomorphisms, which are exact symmetries of
the action of general relativity.  Operators that commute with the
constraints --- for instance, the holonomies of the Chern-Simons
formulation --- are consequently time-independent.  In this context,
the basic problem can be posed quite simply: how does one describe
dynamics when all observables are constants of motion?

Attempts to address this problem, even in very simple models, have been
plagued by ambiguities in operator ordering \cite{Rovelli,Haj}.  As
Kucha{\v r} has stressed \cite{Kuchar}, such ambiguities can hide a
multitude of sins, and no theory should be considered complete unless it
offers a clear ordering prescription.  So far, the only hint of such a
prescription has come from the behavior of the mapping class group, the
group of ``large'' diffeomorphisms.  In $(2+1)$-dimensional gravity with
a sufficiently simple topology, it is known that the requirement of
good behavior under the action of this group places strong restrictions
on possible operator orderings and quantizations \cite{time}.  In this
aspect, $(2+1)$-dimensional quantum gravity resembles two-dimensional
rational conformal field theory, where the representation theory of the
mapping class group plays an important role in limiting the range of
possible models \cite{Moore}.

In this Letter, we shall explore these restrictions in more detail. As we
shall see, even the simplest nontrivial topology, $[0,1]\times T^2$, is
rich enough to illustrate both the importance of the mapping class group
in determining operator orderings and its limitations.

\section{Chern-Simons and ADM Quantization}

A systematic exploration of the relationship between Chern-Simons and
Arnowitt-Deser-Misner (ADM) quantization of $(2+1)$-dimensional gravity
was begun in references \cite{time} and \cite{dirac}.  In this section
we briefly summarize the results; for more details, the reader is referred
to the original papers.\footnote{The notation in this Letter has been
changed slightly to conform to the mathematical literature.  The modulus
of a torus $T^2$, previously denoted by $m$, is now $\tau$, while the trace
of the extrinsic curvature, previously $\tau$, is now $T$.}

In the ADM formulation of canonical $(2+1)$-dimensional gravity
\cite{Moncrief, HosNak}, one begins by specifying a time-slicing.  A
convenient choice is York's ``extrinsic time,'' in which spacetime is
foliated by surfaces of constant mean extrinsic curvature $\hbox{Tr} K
= T$.  For a spacetime with the topology $[0,1]\times T^2$, a slice of
constant $T$ is a torus with an intrinsic geometry that can be characterized
by a complex modulus $\tau = \tau_1 + i\tau_2$ and a conformal factor.
Moncrief has shown that the conformal factor is uniquely determined by
the constraints \cite{Moncrief}, so the physical phase space is
parameterized by $\tau(T)$, its conjugate momentum $\bar p(T)$, and
their complex conjugates.  This behavior is characteristic of
$(2+1)$-dimensional gravity --- all but a finite number of degrees
of freedom are fixed by constraints, and the dynamics takes place on
a finite-dimensional reduced phase space of global geometric variables.
For the torus, in particular, it is not hard to show that the classical
dynamics is determined by the Hamiltonian
\beq
H = T^{-1}\left(\tau_2^{\ 2}\,\bar p p\right)^{1/2} \ .
\label{1x1}
\eeq

In the Chern-Simons formulation \cite{Wita}, in contrast, a classical
solution of the Einstein field equations is characterized by a flat
ISO($2,1$) connection
\beq
A_\mu = e^a_{\ \mu}{\cal P}_a + \omega^a_{\ \mu}{\cal J}_a \ ,
\label{1x2}
\eeq
where $e^a_{\ \mu}$ is the triad, $\omega^a_{\ \mu}$ is the spin connection,
and $\{{\cal P}_a,{\cal J}_a\}$ generate the $(2+1)$-dimensional Poincar\'e
group.  Up to gauge transformations, such a flat connection is completely
determined by its holonomy group $\Gamma\subset\hbox{ISO}(2,1)$.  $\Gamma$
is a group of isometries of the Minkowski metric, and has a simple geometric
interpretation: any classical $(2+1)$-dimensional spacetime is $M$ flat,
and if the topology and causal structure are sufficiently simple,  $M$
may be ``uniformized'' by $\Gamma$ \cite{Mess} --- that is,  $M =
{\cal F}/\Gamma$, where $\cal F$ is some region of Minkowski space on
which $\Gamma$ acts properly discontinuously as a group of isometries.

In particular, if $M$ has the topology $[0,1]\times T^2$, its fundamental
group is $\pi_1(M) = \IZ\oplus\IZ$, so $\Gamma$ is generated by two commuting
Poincar\'e transformations.  In the relevant topological component of the
space of flat connections \cite{Mess}, these can be chosen to be of the
form
\begin{eqnarray}
  \Lambda_1&:& (t,x,y)\rightarrow(t\cosh\lambda+x\sinh\lambda,\,
           x\cosh\lambda+t\sinh\lambda,\,y+a) \nonumber \\
  \Lambda_2&:& (t,x,y)\rightarrow(t\cosh\mu+x\sinh\mu,\,
           x\cosh\mu+t\sinh\mu,\,y+b)\ \ .
\label{1x3}
\end{eqnarray}
A spacetime is thus characterized by four time-independent parameters
$a$, $b$, $\lambda$, and $\mu$, which in a sense already provide a
``frozen time'' picture at the classical level.  Of course, there is
nothing paradoxical here: as in ordinary Hamilton-Jacobi theory, we have
simply described the classical solutions of the field equations in terms
of a set of constants of motion.

For the torus topology, the quotient space ${\cal F}/\langle\Lambda_1,
\Lambda_2\rangle$ can be worked out explicitly and compared to the ADM
metric \cite{time}.  One finds that the ADM moduli at York time $T$ are
\beq
\tau = \Biggl(a+{i\lambda\over T}\Biggr)^{-1}\Biggl(b+{i\mu\over T}\Biggr) \ ,
\label{1x4}
\eeq
and
\beq
p = -iT\left(a-{i\lambda\over T}\right)^2 \ .
\label{1x5}
\eeq
Equations (\ref{1x4}--\ref{1x5}) can be viewed as time-dependent canonical
transformations; the Poisson brackets
\beq
\{a,\mu\} = -\{b,\lambda\} = {1\over2}
\label{1x6}
\eeq
induce corresponding brackets
\beq
\{\tau, \bar p\} = \{\bar\tau,p\} = 2 \ .
\label{1x7}
\eeq
In terms of the holonomy variables, the Hamiltonian \rref{1x1} is
\beq
H = {a\mu - \lambda b \over T} \ ,
\label{1x8}
\eeq
and it is easy to check that the moduli and momenta (\ref{1x4}--\ref{1x5})
obey Hamilton's equations of motion,
\beq
{d\tau\over dT} = -\{H,\tau\} \ ,\qquad\quad {dp\over dT} = - \{H,p\} \ .
\label{1x8a}
\eeq

Note that the relation \rref{1x4} of moduli and holonomies respects
the action of the mapping class group of the torus (also known as the
modular group).  At the level of holonomies, this group is generated by
the two transformations
\begin{eqnarray}
  S&:& (a,\lambda)\rightarrow(b,\mu),\quad
               (b,\mu)\rightarrow(-a,-\lambda)\nonumber \\
  T&:& (a,\lambda)\rightarrow(a,\lambda),\quad
               (b,\mu)\rightarrow(b+a,\mu+\lambda) \ ,
\label{1x9}
\end{eqnarray}
whose form follows in a straighforward manner from the action of the
mapping class group on $\pi_1(T^2)$.  If $\tau$ is defined as in equation
\rref{1x4}, the induced transformations are then
\begin{eqnarray}
  S&:& \tau\rightarrow-{1\over\tau}\nonumber \\
  T&:& \tau\rightarrow\tau+1 \ ,
\label{1x10}
\end{eqnarray}
which may be recognized as the standard generators of the modular group.

It is natural to try to extend these classical relationships to the
quantum theory.  In Chern-Simons quantization, the wave function
$\psi(\lambda,\mu)$ is time-independent, and the absence of dynamics
provides a clear illustration of the ``problem of time.''  One solution
is to interpret the Chern-Simons quantum theory as a Heisenberg
picture, in which wave functions {\em should} be time-independent,
and to look for appropriate time-dependent operators.  In particular,
we might expect the operators representing the ADM moduli $\tau$ and
$\bar\tau$ to depend explicitly on a ``time'' parameter $T$, with
dynamics described by the appropriate Heisenberg equations of motion
corresponding to \rref{1x8a}.  If we pass to a Schr\"odinger picture by
simultaneously diagonalizing $\hat\tau$ and $\hat\tau^\dagger$, we will
obtain wave functions that depend on $T$ as well.  Different choices of
classical time-slicing will lead to different wave functions, of course,
but this may simply reflect the fact that they describe different aspects
of the physics.  This approach to quantization has been advocated by
Rovelli \cite{Rovelli}, who argues that it provides a natural solution
to the problem of time in quantum gravity.

\section{Operator Ordering and the Modular Group}

As might be anticipated, the basic difficulty with such a program is one
of operator ordering.  A classical variable like $\tau$ does not uniquely
determine an operator in the quantum theory: given a candidate operator
$\hat\tau$, any other operator of the form $\hat\tau' = {\hat V}^{-1}
\hat\tau\hat V$ will have the same classical limit.  Of course, if $\psi$
is an eigenfunction of $\hat\tau$, $\psi' = {\hat V}^{-1}\psi$ will be an
eigenfunction of $\hat\tau'$; but unless $\hat V$ happens to be unitary,
$\psi'$ will no longer be a simultaneous eigenfunction of the adjoint
$(\hat\tau')^\dagger$.  We thus run the risk of obtaining not a single
Schr\"odinger picture, but many.

This ambiguity can alternatively be expressed --- at least in this simple
context --- as an ambiguity in the definition of the inner product.
Given an inner product $\langle\psi|\chi\rangle$, we can define a new
$\langle\psi|\chi\rangle' = \langle{\hat V}^{-1}\psi|{\hat V}^{-1}\chi
\rangle$.  The adjoint of $\hat\tau$ then becomes
\beq
\hat\tau^{\dagger\prime} = (\hat V\hat V^\dagger)\hat\tau^\dagger
  (\hat V \hat V^\dagger)^{-1}\ .
\label{2x2a}
\eeq
This observation connects our approach to quantization to the $C^*$-algebra
methods developed in quantum field theory \cite{Haag}; in particular,
the relationship between ADM and Chern-Simons operator algebras found in
\cite{time} is not complete until the choice of adjoint is specified.

For our simple $[0,1]\times T^2$ topology, a natural choice of ordering
is that of equations (\ref{1x4}--\ref{1x5}),
\beq
\hat\tau = \Biggl(\hat a+{i\hat\lambda\over T}\Biggr)^{-1}
  \Biggl(\hat b+{i\hat\mu\over T}\Biggr) \ , \qquad\quad
\hat p = -iT\left(\hat a-{i\hat\lambda\over T}\right)^2 \ .
\label{2x1}
\eeq
It is not hard to check that this ordering preserves the classical
correspondence between modular transformations in the Chern-Simons
and ADM pictures; among simple (rational) orderings, it seems to be
essentially unique in this respect.  Once these definitions have been
chosen, the Hamiltonian is fixed up to an additive constant by the
requirement that the Heisenberg equations of motion hold; it is
\beq
\hat H = {\hat a\hat\mu - \hat\lambda \hat b \over T} \ .
\label{2x2}
\eeq
In reference \cite{dirac}, it was shown that these choices lead to
simultaneous eigenfunctions of $\hat\tau$ and $\hat\tau^\dagger$ that
are automorphic forms of weight $1/2$, that is, spinors on moduli
space.

To generalize this result, let us try to diagonalize ${\hat V}^{-1}\hat\tau
\hat V$ and its adjoint for some arbitrary operator $\hat V$.  Equivalently,
we shall look for the simultaneous eigenfunctions of $\hat\tau$ and the
adjoint $\hat\tau^{\dagger\prime}$ defined in \rref{2x2a}.  As we shall see,
this task is manageable because of the simple form of the eigenfunctions
of $\hat\tau$.

{}From the Poisson brackets \rref{1x6}, we can represent $\hat a$ and $\hat b$
as
\beq
\hat a = {i\over2} {\partial\ \over\partial\mu} \ ,
\qquad\quad \hat b = -{i\over2} {\partial\ \over\partial\lambda} \ ,
\label{2x3}
\eeq
so
\beq
\hat\tau =
  -\left({1\over2}{\partial\ \over\partial\mu} + {\lambda\over T}\right)^{-1}
  \left({1\over2}{\partial\ \over\partial\lambda} - {\mu\over T}\right) \ .
\label{2x4}
\eeq
The general eigenfunction of $\hat\tau$ with eigenvalue $\tau = \tau_1
+ i\tau_2$ is then of the form
\beq
K(\tau,\bar\tau; \lambda,\mu,T)=
  f\left[{\mu - \tau\lambda\over \tau_2^{\ 1/2}T}\right]
   \exp\left\{-{i|\mu - \tau\lambda|^2\over \tau_2 T}\right\} \ ,
\label{2x5}
\eeq
where $f$ is an arbitrary analytic function and the dependence of the
prefactor on $T$ has been chosen so that
\beq
-i{\partial\ \over\partial T}K(\tau,\bar\tau; \lambda,\mu,T)
  = \hat H K(\tau,\bar\tau; \lambda,\mu,T) \ .
\label{2x6}
\eeq
Taylor expanding $f$, we are thus led to consider functions of the form
\beq
K^{(n)}(\tau,\bar\tau; \lambda,\mu,T) =
  \left({\mu - \tau\lambda\over \tau_2^{\ 1/2}T}\right)^n
  \exp\left\{-{i|\mu - \tau\lambda|^2\over \tau_2 T}\right\} \ .
\label{2x7}
\eeq

These functions have a number of useful properties.  First, under the
simultaneous modular transformations \rref{1x9} of $\mu$ and $\lambda$
and \rref{1x10} of $\tau$ and $\bar\tau$, we have
\begin{eqnarray}
  S&:& K^{(n)}\rightarrow \left({\bar\tau\over\tau}\right)^{n/2}K^{(n)}
  \nonumber \\
  T&:& K^{(n)}\rightarrow K^{(n)} \ ,
\label{2x8}
\end{eqnarray}
which means that $K^{(n)}$ is an automorphic form of weight $-n/2$
\cite{Fay,Maass}, essentially a tensor on moduli space.  Second, it is not
hard to show that
\beq
\hat H^2 K^{(n)} =
  T^{-2}\left( \Delta_{-n/2} - {(n-1)^2\over4}\right) K^{(n)} \ ,
\label{2x9}
\eeq
where $\hat H$ is the Hamiltonian \rref{2x2} and
\beq
\Delta_k = -\tau_2^{\ 2}\left( {\partial^2\ \over\partial\tau_1^{\ 2}} +
  {\partial^2\ \over\partial\tau_2^{\ 2}}\right)
  + 2ik\tau_2{\partial\ \over\partial\tau_1} + k(k+1) \
\label{2x10}
\eeq
is the Maass Laplacian acting on automorphic forms of weight $k$.
Together, these characteristics allow us to employ a large body of
mathematical work on representations of the modular group; for instance,
the Maass Laplacian is simply an SL($2,\IR$) Casimir operator \cite{Fay}.

Let us now expand a Chern-Simons wave function $\psi(\mu,\lambda)$ in terms
of the eigenfunctions $K^{(n)}(\tau,\bar\tau; \lambda,\mu,T)$:
\beq
\psi(\mu,\lambda) = \int {d^2\tau\over\tau_2^{\ 2}}\,
  K^{(n)}(\tau,\bar\tau; \lambda,\mu,T)\tilde\psi^{(n)}(\tau,\bar\tau,T) \ .
\label{2x11}
\eeq
For $\psi(\mu,\lambda)$ to be invariant under modular transformations,
$\tilde\psi^{(n)}(\tau,\bar\tau,T)$ must transform with a phase that
cancels that of $K^{(n)}(\tau,\bar\tau; \lambda,\mu,T)$, i.e., it must be
an automorphic form of weight $n/2$.  Moreover, since $\psi(\mu,\lambda)$
is independent of $T$, equation \rref{2x6} gives
\begin{eqnarray}
0 = \left( T{\partial\ \over\partial T}\right)^2\psi(\mu,\lambda)
  = \int {d^2\tau\over\tau_2^{\ 2}}\Biggl[ &-&T^2\left(\hat H^2 K^{(n)}
  (\tau,\bar\tau; \lambda,\mu,T)\right)\tilde\psi^{(n)}(\tau,\bar\tau,T)
  \nonumber \\
  &+& K^{(n)}(\tau,\bar\tau; \lambda,\mu,T)
  \left( T{\partial\ \over\partial T}\right)^{\!2}
  \tilde\psi^{(n)}(\tau,\bar\tau,T) \Biggr] \ .
\label{2x12}
\end{eqnarray}
Using \rref{2x9} and integrating by parts, we find that
\beq
\left( T{\partial\ \over\partial T}\right)^2
  \tilde\psi^{(n)}(\tau,\bar\tau,T) =
  \left[\Delta_{n/2}-{(n+1)^2\over4}\right]\tilde\psi^{(n)}(\tau,\bar\tau,T)
  \ ,
\label{2x13}
\eeq
which is the Klein-Gordon equation for an automorphic form of weight
$n/2$.  Our diagonalization procedure thus produces wave functions that
obey a simple wave equation of the type one might expect from direct ADM
quantization.

For $n=1$, this analysis reduces to that of \cite{dirac}.  In particular,
it is not hard to check that
\beq
\hat\tau^\dagger K^{(1)}(\tau,\bar\tau; \lambda,\mu,T) =
  \bar\tau K^{(1)}(\tau,\bar\tau; \lambda,\mu,T) \ ,
\label{2x14}
\eeq
so the diagonalization of $\hat\tau$ and $\hat\tau^\dagger$ with the
ordering \rref{2x1} leads to ADM-type wave functions that transform as
forms of weight $1/2$.  Naive ADM quantization, on the other hand, gives
modular invariant wave functions, i.e., forms of weight $0$ \cite{HosNak}.
To understand such functions, observe that
\beq
\hat\tau^* K^{(0)}(\tau,\bar\tau; \lambda,\mu,T) =
  \Biggl(\hat a-{i\hat\lambda\over T}\Biggr)^{-1}
  \Biggl(\hat b-{i\hat\mu\over T}\Biggr)
  K^{(0)}(\tau,\bar\tau; \lambda,\mu,T) =
  \bar\tau K^{(0)}(\tau,\bar\tau; \lambda,\mu,T) \ ,
\label{2x15}
\eeq
so $n=0$ corresponds to the simultaneous diagionalization of $\hat\tau$
and $\hat\tau^*$.  Moreover, it is easy to check that
\beq
\hat\tau^* =
  (\hat p\hat p^\dagger)^{-1/2}\hat\tau^\dagger(\hat p\hat p^\dagger)^{1/2}
  \ .
\label{2x16}
\eeq
Standard ADM quantum theory thus corresponds to the operator ordering
\rref{2x2a} with $\hat V = \hat p^{-1/2}$.  Other orderings will give rise
to automorphic forms with other weights, but at least for orderings of the
form $\hat\tau' = \hat V^{-1}\hat\tau\hat V$, we have now seen that this
is the only ambiguity.

\section{Implications}

Kucha{\v r} has described the approach to quantization taken here as one
of ``evolving constants of motion'' \cite{Kuchar}.  He points out a number
of potential problems, including operator ordering ambiguities and possible
dependence on the choice of time-slicing.\footnote{More precisely, he
suggests that the dynamical operators corresponding to different choices
of time-slicing may not all be self-adjoint under any choice of inner
product.}

The York time-slicing is the only choice that has been studied in any
detail in $(2+1)$-dimensional gravity, so we cannot yet address the latter
issue.  As for the former, we have seen that operator ordering ambiguities
do indeed make the quantum theory nonunique.  But the range of possible
theories is surprisingly limited, and is essentially determined by the
representation theory of the modular group.  The modular group  thus
plays a role roughly analogous to that of the Poincar\'e group in ordinary
free quantum field theory, determining a small family of admissible models.

Of course, the topology $[0,1]\times T^2$ is exceptionally simple even
for $2+1$ dimensions, and one must worry about whether these results can
be extended to less trivial spacetimes.  In general, the Hamiltonian
\rref{1x1} will become much more complicated --- it will typically
be nonpolynomial --- and it will be difficult to express the ADM moduli
explicitly in terms of holonomies.  It is possible that useful results
can be obtained for spaces of genus two, for which the hyperelliptic
representation provides a major simplification.  In any case, we can at
least be confident that the representation theory of the mapping class
group will continue to play a key role.

\begin{flushleft}
\large\bf Acknowledgements
\end{flushleft}

I would like to thank Arley Anderson for useful discussions of operator
orderings.  This work was supported in part by the Department of Energy
under grant DE-FG03-91ER40674.


\begin{thebibliography}{99}

\bibitem{Rovelli} C.\ Rovelli, Phys.\ Rev. {\bf D42} (1990) 2638;
  Phys.\ Rev. {\bf D43} (1991) 442; Phys.\ Rev. {\bf D44} (1991) 1339.
\bibitem{Haj} P.\ H\'aj\'\i\v cek, Phys.\ Rev. {\bf D44} (1991) 1337.
\bibitem{Kuchar} K.\ Kucha{\v r}, in Proc.\ of the 4th Canadian Conference
  on General Relativity and Relativistic Astrophysics, edited by G.\
  Kunstatter et al.\ (World Scientific, Singapore, 1992).
\bibitem{time} S.\ Carlip, Phys.\ Rev.\ {\bf D42} (1990) 2647.
\bibitem{Moore} G.\ Moore and N.\ Seiberg, in Superstrings '89 (World
  Scientific, Singapore, 1990).
\bibitem{dirac} S.\ Carlip, Phys.\ Rev.\ {\bf D45} (1992) 3584.
\bibitem{Moncrief} V.\ Moncrief, J.~Math.\ Phys.\ {\bf 30} (1989) 2907.
\bibitem{HosNak} A.\ Hosoya and K.\ Nakao, Prog.\ Theor.\ Phys.\ {\bf 84}
  (1990) 739.
\bibitem{Wita} E.\ Witten, Nucl.\ Phys.\ {\bf B311} (1988) 46.
\bibitem{Mess} G.~Mess, Institut des Hautes Estudes Scientifiques
  preprint IHES/M/90/28 (1990).
\bibitem{Haag} R.\ Haag and D.\ Kastler, J.~Math.\ Phys.\ {\bf 5} (1964)
  848.
\bibitem{Fay} J.~D.~Fay, J.~Reine\ Angew.\ Math.\ {\bf 293} (1977) 143.
\bibitem{Maass} H.\ Maass, Lectures on Modular Functions of One Complex
  Variable (Tata Institute, Bombay, 1964).

\end{thebibliography}
\end{document}